\documentclass[a4paper]{article}

\usepackage{INTERSPEECH2022}
\usepackage{multirow}
\usepackage{soul}
\usepackage{xcolor}

\usepackage[font=footnotesize,labelfont=bf]{caption}
\usepackage{eso-pic, rotating, graphicx}

\title{\vspace{-1mm}Federated Self-supervised Speech Representations: Are We There Yet?}
\name{Yan Gao$^1$, Javier Fernandez-Marques$^2$, Titouan Parcollet$^3$, Abhinav Mehrotra$^2$, Nicholas D. Lane$^{1,2}$}

\address{\vspace{-1mm}
  $^1$University of Cambridge, $^2$Samsung AI, $^3$Avignon University}
  
\begin{document}

\maketitle
\begin{abstract}
The ubiquity of microphone-enabled devices has lead to large amounts of unlabelled audio data being produced at the edge. The integration of self-supervised learning (SSL) and federated learning (FL) into one coherent system can potentially offer data privacy guarantees while also advancing the quality and robustness of speech representations. 
In this paper, we provide a \textit{first-of-its-kind} systematic study of the feasibility and complexities for training speech SSL models under FL scenarios from the perspective of algorithms, hardware, and systems limits. 
Despite the high potential of their combination, we find existing system constraints and algorithmic behaviour make SSL and FL systems nearly impossible to build today. Yet critically, our results indicate specific performance bottlenecks and research opportunities that would allow this situation to be reversed. While our analysis suggests that, given existing trends in hardware, hybrid SSL and FL speech systems will not be viable until 2027, we believe this study can act as a roadmap to accelerate work towards reaching this milestone much earlier.

\end{abstract}

\section{Introduction}
A large amount of audio data, often unlabelled and of private nature, is generated everyday from cellphones, tablets, personal assistants and other IoT devices \cite{arakawa2019implementation,nguyen2012pattern,maloney2001can,shah2018audio}. Being able to utilise this data to solve various speech related tasks has been of great interest to researchers for over a decade \cite{baevski2020wav2vec,schneider2019wav2vec,liu2020mockingjay,ravanelli2020multi,pascual2019learning}. Self-supervised learning (SSL) allows the learning of representations from unlabelled data, which can later be used to solve specific downstream tasks, e.g., automatic speech recognition (ASR), speech translation, keyword spotting (KWS), and others. SSL shares limitations with other forms of unsupervised training favouring large-capacity models trained with vast amounts of data \cite{baevski2020wav2vec,schneider2019wav2vec}. Currently, such workloads can only be realised in data centers, where high-performance hardware is available. Such centralised SSL systems unavoidably raise concerns around privacy specially when dealing with speech data. Additionally, they also cause severe communication overheads due to the transferring of data from devices where it originates. 

\begin{figure}[t]
  \centering
  \includegraphics[width=0.98\linewidth]{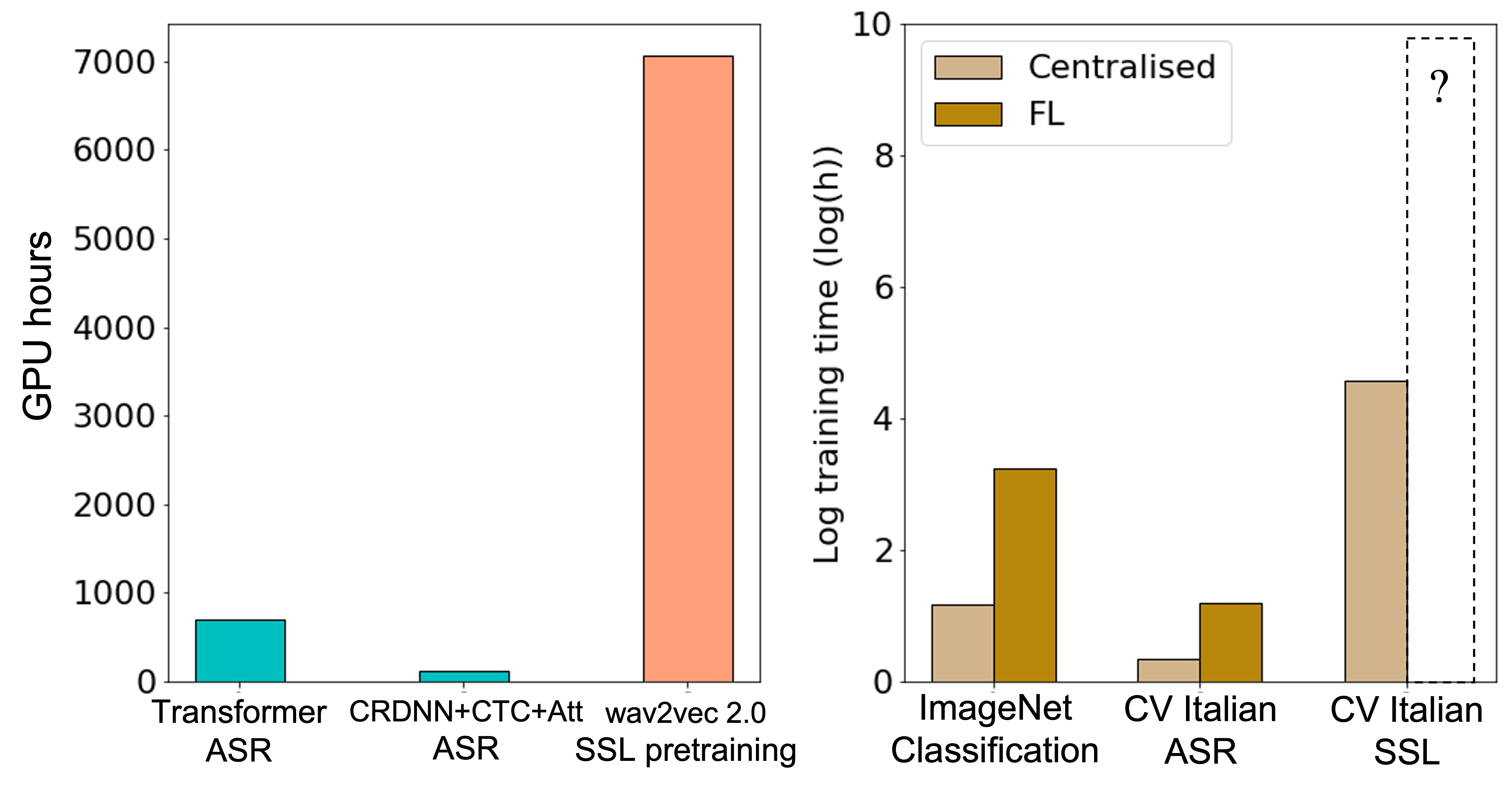}
    \vspace{-0.3cm}
  \caption{\small (left) The pretraining of a SOTA self-supervised model wav2vec 2.0 is several times more time-consuming than SOTA supervised ASR models (transformer \cite{vaswani2017attention,speechbrain}, CRDNN+CTC+Att \cite{kim2017joint,speechbrain}). All of the models are trained with the LibriSpeech dataset using V100 GPUs. (right) FL leads to longer training time compared to centralised settings. 
  The models are trained using V100 GPU in centralised settings and using Jetson Xavier NX in FL \cite{qiu2021first}. CV is Common Voice dataset.
  }
  \label{fig:1}
  \vspace{-0.5cm}
\end{figure}

A natural way to mitigate these issues is to combine SSL with federated learning (FL) \cite{mcmahan2017communication,gao2021end,leroy2019federated,granqvist2020improving}.
In FL, a distributed population of devices collaboratively trains a single global model while keeping their local data private. A typical FL pipeline is comprised of three stages: first, a global model is initialised in a server to which all FL nodes (or \textit{clients}) connect; then, a fraction of clients are chosen by the server, receive a copy of the global model and each performs on-device training using their own data; upon completion the clients send the updated model to the server, where a new global model is generated by aggregating the client updates. This process is repeated for a number of rounds. 
Deploying SSL workloads in a federated setting offers another benefit beyond privacy-preserving training: large-scale distributed feature learning from real-world data without costly and error-prone manual annotations. 
\textit{Despite the combination offering significant benefits, no prior work has studied speech SSL in the context of FL.} 

The SSL models are notoriously costly to train.
A SOTA SSL model for speech representation learning, wav2vec 2.0 \cite{baevski2020wav2vec}, requires over $7000$ GPU hours of pretraining (V100 GPUs). That is an order of magnitude more compute resources than a top-performing model trained with only supervision (Fig. \ref{fig:1} left). However, this leap in resource demands when transitioning from supervised to a SSL solution is not the only significant barrier to performing SSL under FL. It is paired with an empirical phenomena generally seen when migrating from centralised training to a federated version: due to the model update aggregation phase of FL, the overall global convergence is weakened and thus slower (Fig. \ref{fig:1} right). We again observe resource requirements (in this case raw training time) increasingly alarmingly -- although this is now due to the act of federating the training process. While we know these two factors will drive the ultimate performance of systems that combine FL and SSL for speech, curiously, there is large gap in existing empirical knowledge as to how they will interact and translate into resource requirements. How long would a SSL speech model take to converge if federated? Specifically, what would the memory and compute requirements be for the participating FL clients? Would WER decrease in this decentralised setting? \textit{All of these questions, ranging from those that are obvious down to questions specific in detail are virtually unknown presently. } 
%
%
%
%

\begin{figure*}[t]
  \centering
  \includegraphics[width=\linewidth]{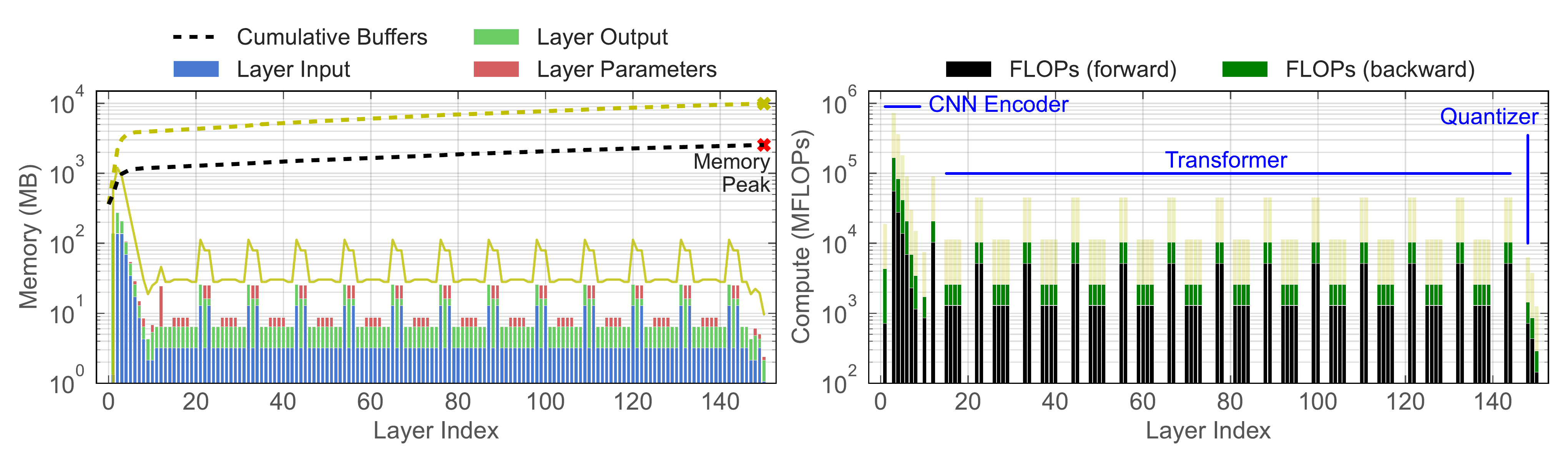}
  \vspace{-7mm}
  \caption{\small Per-layer breakdown of wav2vec 2.0 \textit{base} model in terms of its memory consumption (left) and compute footprint (right). The data referred to in the legend field is obtained when training with inputs of length 5.5s (the average in CV Italian), batch size 4 and resampled at 16KHz. (Left) The CNN encoder block contributes to a significant memory utilization compared to the transformer layers that follow. At the end of the forward pass the memory utilization is at its maximum (shown with a red cross) since intermediate layer activations need to be kept in-memory in order to do backpropagation. 
  In yellow we show the memory utilization for the same model but when using sequences of 12s and batch size 8. The solid yellow line shows the per-layer total memory utilization whereas the dashed one shows the cumulative memory needed for training. In this scenario training requires just under 10 GB of memory on top the memory usage of other processes responsible for data loading, checkpointing and the OS itself. (Right) Per-layer breakdown of the amount of compute (FLOPs) needed for training. At the back, yellow bars represent the total FLOPs when using with sequences of 12s and batch size 8. Early layers in the CNN block require over an order of magnitude more compute.}
  \label{fig:layers}
  \vspace{-0.3cm}
\end{figure*}


In this work, we conduct a systematic investigation as to the precise bottlenecks that cause SSL models to be so resource intensive to train, and further explore how these resource requirements (e.g., memory, compute) translate into the feasibility of SSL models on edge devices in FL environments. 
We find that the modest edge devices and existing training frameworks are not yet a match for the requirements of federating SSL speech. But more crucially, our empirical results provide for-the-first-time the much needed identification and quantification of the key limiting factors that underpin this situation. As an example of how our findings can accelerate research in this emerging area, we use them to propose a novel modification to FL aggregation that takes into account the training loss of individual FL clients. Our study estimates given existing FL and SSL technique and practices -- along with historically guided expectation of hardware improvements~\cite{schaller1997moore,sevilla2022compute} -- that hybrid self-supervised FL for speech will not be feasible until roughly half decade from now, 2027; even though the core principles are available today. Not only is this the first systematic estimate of its type for the speech community to reflect upon, we hope that it acts as a call to arms around which researchers will rally in an effort to reach this milestone much more quickly. 
%
%


The scientific contributions of this study are: (1) We conduct a systematic analysis that discovers the core reasons why SSL models are resource intensive (\S\ref{sec:model} and \S\ref{sec:length}); (2) We extend this analysis to explore the feasibility of SSL under FL conditions (\S\ref{sec:fea}); (3) We perform extrapolative analysis as to future hardware feasibility through faithful GPU simulations (\S\ref{sec:extra}). 

\section{Experimental Settings}
\label{sec:design}
We conduct all of the measurements using wav2vec 2.0 \cite{baevski2020wav2vec} on the CommonVoice Italian (CV Italian) dataset \cite{ardila2019common} trained on a variety of hardware platforms. The details are as follows:
\vspace{1mm}

\noindent \textbf{Model.} Our choice of SOTA SSL model for speech representation learning is wav2vec 2.0. This model is comprised of three modules: a convolution layers block that extract latent acoustic features from the raw waveform; a series of transformer layers that generate contextualised embeddings; and, a quantization module that discretises the output of the convolutional module to a finite set of speech representations. The model is jointly trained by a contrastive loss and a diversity loss. We consider both the \textit{base} and \textit{large} versions of wav2vec in our evaluation.
\vspace{1mm}

\noindent \textbf{Dataset.} The Common Voice Italian corpus (version 7.0) contains a total of $211$K utterances ($317$ hours) obtained from volunteers recording sentences on smartphones, computers, tablets, etc. The validated set is used for SSL pretraining and consists of $195$K utterances ($288$ hours) with an average sentence length of $5.5$ seconds from more than $6$K speakers. The training set is used for downstream ASR fine-tuning which contains $191$ hours of speech, while the validation and test sets contain around $25$ hours of speech. 
Due to their heterogeneity in terms of recordings and speaker diversity, CV datasets are seen as a perfect match to conduct FL-related experiments \cite{gao2021end}.
\vspace{1mm}

\noindent \textbf{Devices.} Our study considers different classes of devices for training speech SSL models. Server-grade NVIDIA A40 GPUs with 48GB of VRAM are used to represent centralised training, while MacBook Pro 2019, Raspberry Pi 4 and NVIDIA Jetson Xavier NX and AGX represent four tiers of FL devices. We also consider different mobile processors for memory comparison, including Snapdragon 870, A15 Bionic and Google Tensor.

\begin{table}[t]
\begin{center}
    \caption{\small High-level breakdown of the number of parameters and FLOPs for inference when considering a sequence of $5.5$ seconds for three main modules in wav2vec 2.0.}
    \vspace{-0.3cm}
    \label{tab:model_size}
    \scalebox{0.9}{
    \begin{tabular}{l rrrr}
    \toprule
      & \multicolumn{2}{c}{Parameters (M)} & \multicolumn{2}{c}{GFLOPs} \\
   \cmidrule(r){2-3} \cmidrule(r){4-5}
    \textbf{Modules} &  \textit{Base} & \textit{Large} & \textit{Base} & \textit{Large} \\
     \cmidrule(r){1-5}
      CNN Encoder & 4.60 & 4.73 & 27.20  & 27.28  \\
      Transformer & 89.78 & 310.70 & 49.16 & 170.16 \\
      Quantization & 0.41 & 0.57 & 0.32 & 0.94 \\
       \cmidrule(r){1-5}
       Total & 94.79 & 316.00 & 76.68 & 198.32 \\
     \bottomrule
    \end{tabular}
    }
\end{center}
\vspace{-0.7cm}
\end{table}

\section{SSL Speech System Resource Analysis}
\label{sec:analysis}

This section provides detailed analysis on the resource footprint of training wav2vec 2.0 (\S\ref{sec:model}), and the sensitivity of resources to input audio length (\S\ref{sec:length}). 
These results then enable a feasibility study of federated SSL under existing hardware (\S\ref{sec:fea}).

\subsection{Architecture breakdown}
\label{sec:model}
Table \ref{tab:model_size} shows the number of parameters and floating operation points (FLOPs) for each of the three modules in wav2vec 2.0 as well as the total model size and FLOPs required for inference while considering an input audio of duration $5.5$ seconds corresponding to the average length of the CV Italian dataset. An immediate observation is the asymmetry between the amount of parameters and FLOPs between the Encoder and Transformer blocks: while the latter has $19\times$ more parameters, it only results in $1.8\times$ more FLOPS compared to the CNN Encoder. When considering the \textit{large} wav2vec 2.0 model, this difference increases significantly: the Transformer has $66\times$ more parameters but only accounts for $6\times$ more FLOPS than the Encoder. This first-order analysis does not consider the training dynamics and therefore a more in-depth look into the impact of each sub-architectural module is needed to identify bottlenecks, both in terms of memory and compute, occurring at training time. 

Figure \ref{fig:layers} shows the per-layer memory consumption (left) and compute (right) of wav2vec 2.0 \textit{base}. Immediately, the high memory utilization and compute needed for the CNN Encoder stands out compared to the rest of the model. These two dimensions are a function of the size of the input tensor passed through the layer -- which is not captured in high-level analysis as in Table \ref{tab:model_size}. When it comes to extending SSL models to constrained devices for FL, a key factor is the total amount of memory needed at training time. Figure \ref{fig:layers} (left) shows a memory peak (red cross) of 2.54GB when training with input sequences of length 5.5s (avg. length) and batch size 4. When considering more desirable batch size of 8 and audio files of 12s, the total memory accumulated prior to backpropagation is 9.89GB. Such a delta between these training settings highlights how seemingly small changes can translate into having to consider an entirely new tier of devices to support such workloads. 


\begin{figure}[t]
  \centering
  \includegraphics[width=0.95\linewidth]{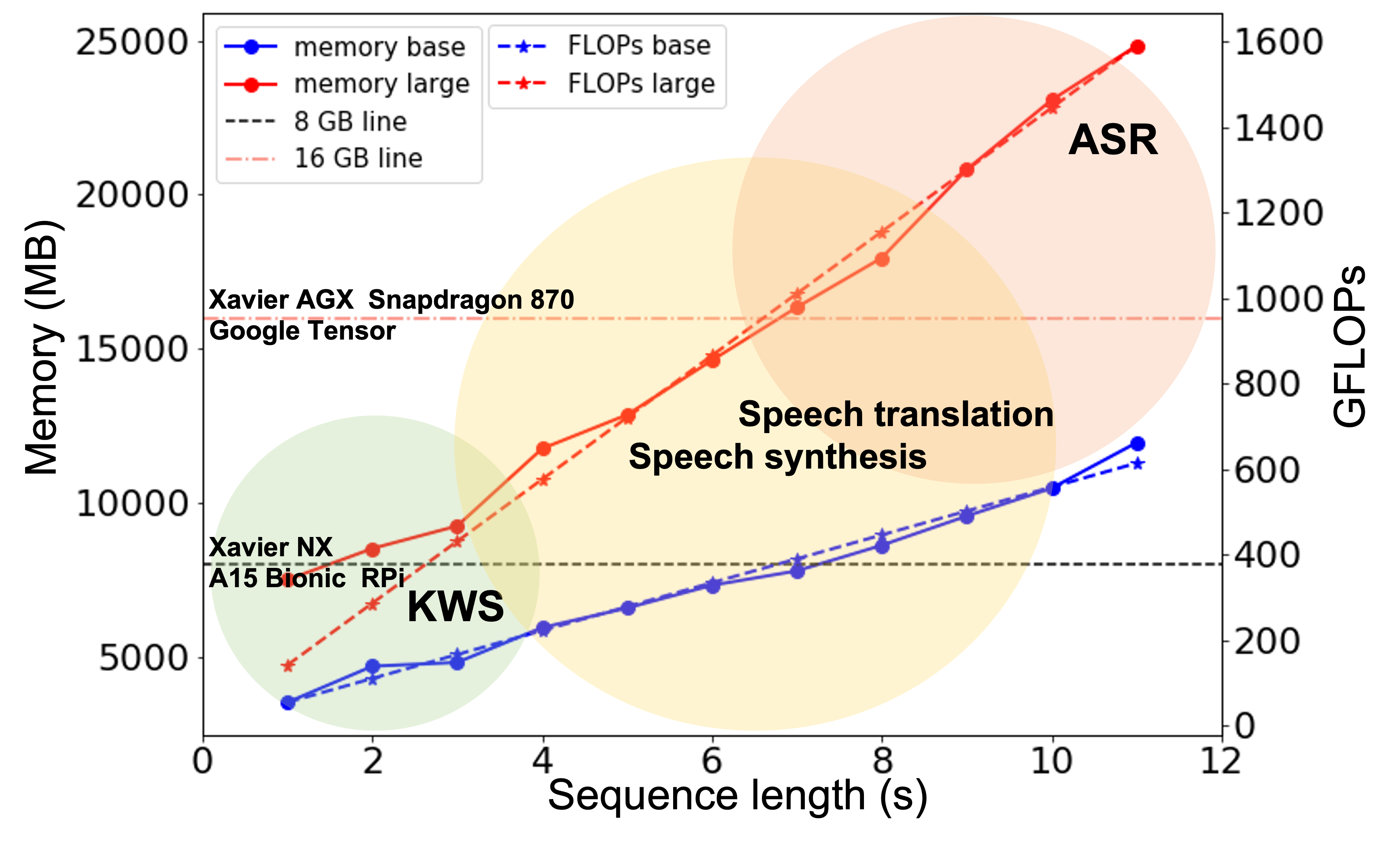}
  \vspace{-3mm}
  \caption{\small Memory utilization and FLOPs with sequence length mapped to relevant offline speech applications and batch of 4.}
  \label{fig:mem_line}
  \vspace{-0.5cm}
\end{figure}

\subsection{On sequence lengths and numerical precision}
\label{sec:length}
Memory and FLOPs grow linearly with the length of the input sequence (Fig. \ref{fig:mem_line}). From a practical perspective, the sequence length can link to different types of offline speech applications that could be regarded as downstream tasks for the pre-trained SSL model. For instance, KWS may only require short utterances (e.g. 1-3s) \cite{chen2014small,zhang2017hello}, while tasks such as speech synthesis or recognition may increase this to much longer sentences \cite{tan2021survey,wang2019overview}. However, currently released large-scale SSL models including wav2vec 2.0, HuBERT \cite{hsu2021hubert} or WavLM \cite{chen2021wavlm} are often trained with sentences longer than 15s, making them clearly intractable for current constrained devices. Hence, one could argue that the SSL stage could be tailored to a specific subset of downstream tasks and limit the length of the input sentences for pretraining, reducing the compute and memory footprints.

\begin{figure}[t]
  \centering
  \vspace{-4mm}
  \includegraphics[width=0.95\linewidth]{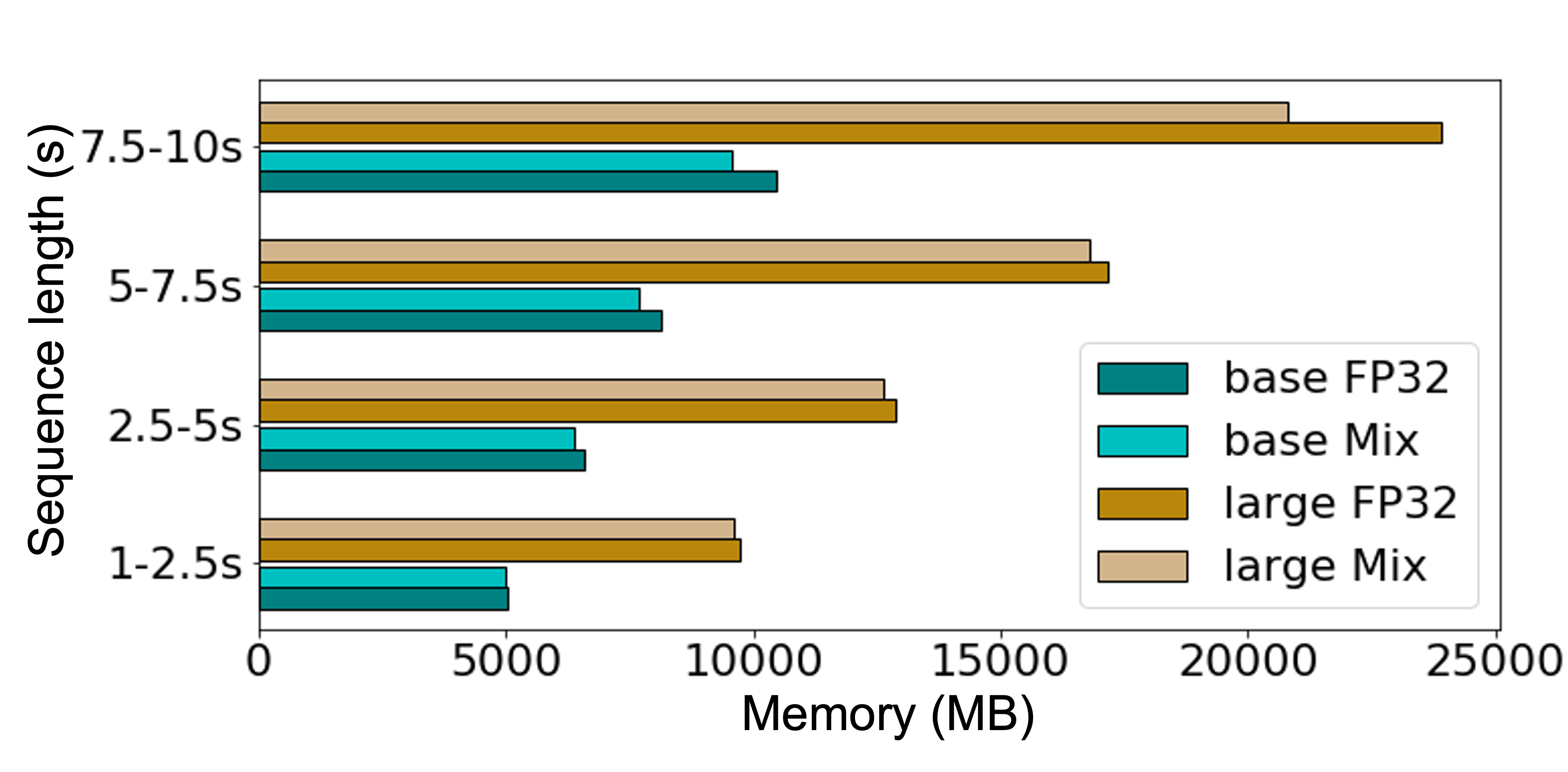}
  \vspace{-0.2cm}
  \caption{\small Memory usage for different sequences length with FP32 and mixed precision (FP32\&FP16) and batch size of 4.
  }
  \label{fig:mem_bar}
  \vspace{-0.5cm}
\end{figure}

Parameter precision is another element that affects the memory and compute footprint of an architecture. Therefore, we compare the memory impact of full-precision \textit{FP32} with mixed precision (\textit{FP32}\&\textit{FP16}) \cite{micikevicius2018mixed} (Fig. \ref{fig:mem_bar}). From the latter results, it appears that the memory savings observed with mixed-precision is not significant and might not be critical while selecting a pool of devices to train SSL models. The impact on the training time, however, provides a different outcome (\S\ref{sec:fea}). 

\subsection{Federated SSL feasibility}
\label{sec:fea}

With this section, we provide an analysis relying on memory consumption and training time answering the feasibility of applying federated learning to large-scale SSL pretraining.
\vspace{1mm}

\noindent \textbf{Memory cost}. 
Memory is a scarce resource in edge devices and limits the feasibility of on-device training in FL. 
For instance, some popular mobile processors and edge devices including the Raspberry Pi 4 (RPi), Jetson Xavier NX or the A15 Bionic chip are only accompanied with $8$GB of memory while the Jetson Xavier AGX, Google Tensor and Snapdragon 870 chips commonly are given $16$GB. According to our findings, the first batch of devices could train \textit{base} wav2vec 2.0 models while the latest set could use \textit{large} architectures with sequences of around $7$ to $8$ seconds at most (Fig. \ref{fig:mem_line}). However, if longer sentences are considered as it is typically the case with current SSL models, all the devices may simply be able to train the \textit{base} model, or even none of them. Nevertheless, we should note that these devices, specially when it comes to smartphones, would likely be running other applications and processes in the background. As a result, the memory ceiling available for FL workloads is lower than the devices' total memory. Therefore, it appears highly probable that even the \textit{base} wav2vec 2.0 architecture exhibits a training memory cost already higher than what high-end devices have to offer.
\vspace{1mm}


\noindent \textbf{Training time}. Training time is another factor influencing the feasibility of SSL in FL. Typical large-scale SSL pretraining takes weeks or months even with hundreds of GPUs \cite{babu2021xls}. Training time might simply become intractable with limited compute devices and the slower convergence induced by FL. 

Hence, we benchmark the observed training time (i.e. seconds per batch) with different FL devices and a modern GPU. As in  prior experiments, sentences are sampled so that the average length is $5.5$s. Table \ref{tab:time} reports the obtained measurements. When considering an unlikely yet simple batch size equal to 1, the 8-cores i9 of the MacBook Pro is $30.3\times$ and $39.5\times$ slower than a Nvidia A40 GPU when training on wav2vec 2.0 \textit{base} and \textit{large} respectively. In the case of the more modest 4-cores RPi, training $4.4\times$ slower than the MacBook Pro and $138\times$ slower then the A40.  The Jetson boards on the other hand are equipped with embedded GPUs and therefore decrease this training time factor to $2.7\times$ and $5.6\times$ slower compared to the A40 for AGX and NX respectively on the \textit{base} model. Once we increase the batch size to 4, which remains considerably low for SSL training that often rely on large batch size values (above 256 \cite{babu2021xls}), the \textit{per-sequence} training times get reduced by 15\%, 20\%, 29\% and 33\% for the MacBook Pro, RPi, AGX and NX respectively thanks to a better hardware utilization. 
Finally, and conversely to the memory cost, 
mixed precision provides a substantial boost in training speed for devices supporting it: $1.56\times$ and $1.31\times$ faster than full precision for the Xavier NX and AGX with the \textit{base} model. However, support for mixed precision training is currently rare in embedded devices. 
%

\begin{table}[t]
\begin{center}
    \caption{\small Training time (average seconds per batch) of wav2vec 2.0 models using sequences with average length $5.5$ seconds. 
    OOM stands for out of memory.}
    \vspace{-0.3cm}
    \label{tab:time}
    \scalebox{0.85}{
    \begin{tabular}{lccccc}
    \toprule
     & & \multicolumn{2}{c}{\textit{Base}} & \multicolumn{2}{c}{\textit{Large}} \\
   \cmidrule(r){3-4} \cmidrule(r){5-6}
    \textbf{Device} & Batch &  FP32 & Mixed & FP32 & Mixed \\
     \cmidrule(r){1-6}
      \multirow{2}{*}{NVIDIA A40} & 1 & 0.12 & 0.11 & 0.23  & 0.21  \\
       & 4 & 0.27 & 0.21 & 0.43 & 0.42  \\
      \cmidrule(r){1-6}
       \multirow{2}{*}{MacBook Pro 2019} & 1 & 3.76  & — & 9.05 & —  \\
      & 4 & 12.83  & — & 33.66 & —  \\
      \midrule
        \multirow{2}{*}{Raspberry Pi 4} & 1 & 16.60 & — & OOM & — \\
      & 4 & 53.26  & — & OOM & —  \\
      \midrule
      \multirow{2}{*}{Jetson Xavier AGX} & 1 & 0.38 & 0.43 & 0.88 & 0.87 \\
      & 4 & 1.08 & 0.82 & OOM & 1.72  \\
      \midrule
      \multirow{2}{*}{Jetson Xavier NX} & 1 & 0.67 & 0.61 & OOM & OOM \\
      & 4 & 1.78  & 1.14 & OOM & OOM  \\
     \bottomrule
    \end{tabular}
    }
\end{center}
\vspace{-0.8cm}
\end{table}

In the more favorable scenario (viz. a A40 GPU and \textit{base} model), we measure the total training time of one epoch for a FL setting with 10 clients ($19.5$K samples per client) to be $0.37$ hours. This value is multiplied by the necessary $150$ epochs (see \S\ref{sec:extra}) of the full FL procedure, hence reaching $2.31$ days, or $55.5$ hours of training. Extending this to our set of FL devices leads us to $110$, $456$, $9$ and $15$ days of training with the MacBook Pro, RPi, AGX and NX respectively i.e., considering that each client trains for the same amount of data concurrently. In practice, the total number of training rounds might be much higher, as it tends to increase with the number of available clients for FL (\S\ref{sec:extra}). Additionally, CV dataset is rather a small dataset for training SSL systems as it solely contains $290$ hours of speech compared to the 10s of thousands of hours of recent pre-trained models. Consequently, and according to our analysis that highlights a training period of already two weeks for a small model on a small dataset, we claim that current experimental protocols leading to large-scale SSL models reaching state-of-the-art downstream performance are simply impractical for the current state of the real-world FL environment.

\section{Federated SSL Speech Representations}
\label{sec:extra}

With this last section, we estimate when FL/SSL speech systems will become feasible for edge hardware; before showing results that \textit{to-the-best-of-our-knowledge} are the first attempt at training SOTA SSL speech under a simulated FL setting. 
%

\subsection{Feasibility under existing hardware trends}
Compute capabilities 
have been doubling every 18 months for decades~\cite{schaller1997moore,sevilla2022compute,hooker2021hardware}. This implies a $8\times$ acceleration of compute in $5$ years. 
Based on our precise measurement of training time on the NX, which represents a typical FL device with an embedded GPU, we predict edge devices to exhibit the same compute performance than a A40 GPU by 2027. 
Thus, assuming the existing SSL models (i.e., no substantial growth in model complexity) and FL methodologies, we must wait for around 5 years for hybrid speech systems combining both technologies to become feasible. 
As discussed in \S\ref{sec:fea}, a large amount of memory is needed when considering longer input sequences. Currently, the AGX can be purchased with 32GB of memory, therefore we believe that by 2027 mid-tier edge systems (e.g. NX) would come with memory sizes similar to today's A40 48GB. This will support performing SSL suitable to more challenging downstream tasks (e.g. ASR).


\begin{figure}[t]
  \centering
  \includegraphics[width=0.9\linewidth]{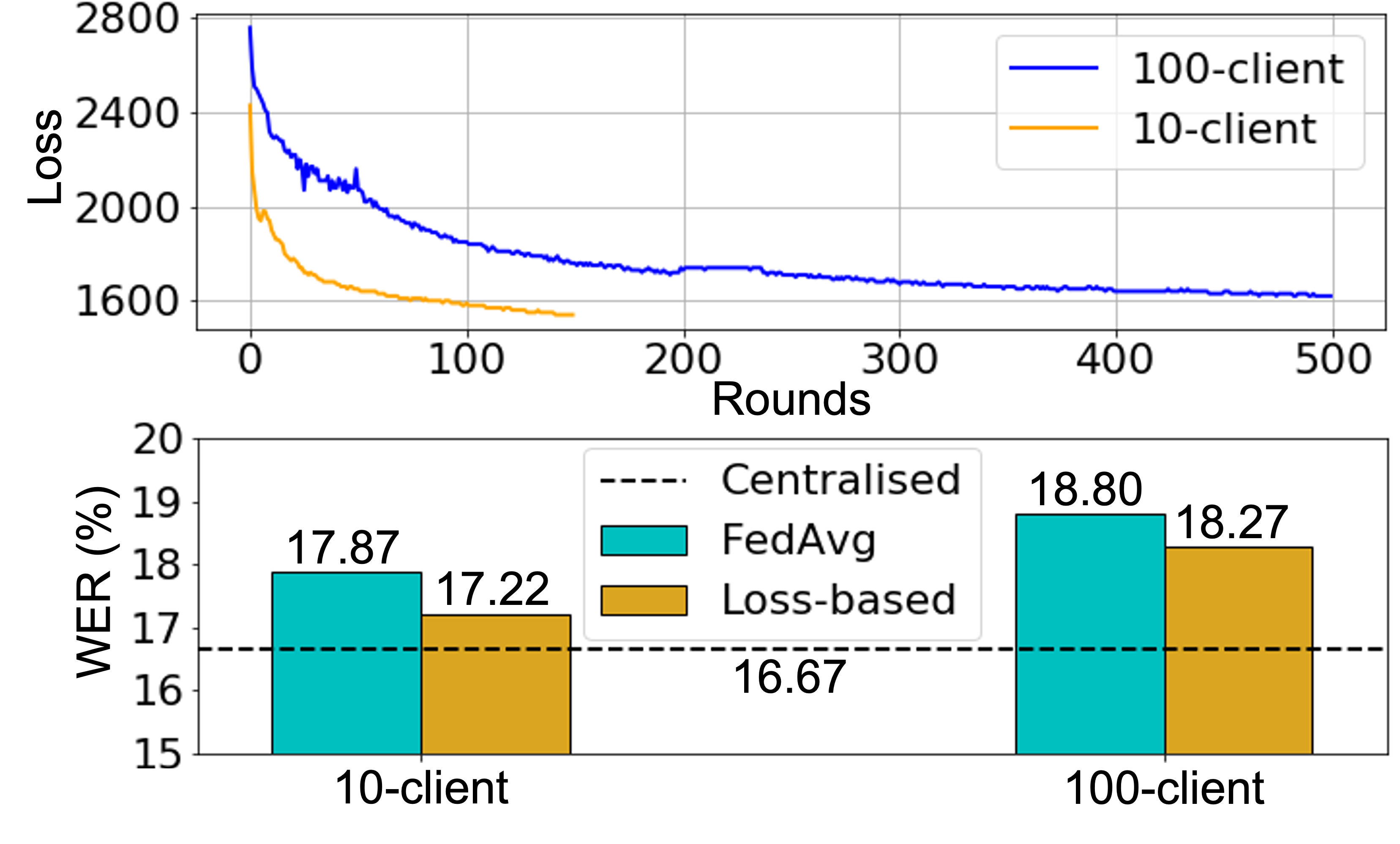}
  \vspace{-2mm}
  \caption{\small (above) 
  Validation loss for SSL pretraining with loss-based aggregation.
  (below) ASR downstream results on test set with SSL representations from FL and centralised settings.}
  \label{fig:loss}
  \vspace{-5mm}
\end{figure}

\subsection{Simulated Federated SSL for Speech}

We train federated wav2vec 2.0 \textit{base} using SpeechBrain \cite{speechbrain} and Flower \cite{beutel2020flower} on two NVIDIA A40 GPUs.

\vspace{1mm}

\noindent \textbf{FL partitions.} Based on the natural partitioning of CV Italian we have two sets of experiments: (1) \textit{10-client FL} and (2) \textit{100-client FL}. In (1), we simulate a simple FL scenario where clients are generally few but with high availability during all rounds, and similar data distribution \cite{kairouz2021advances}. The dataset is randomly split into 10 partitions (i.e. one per client) without overlapping of speakers, each containing roughly the same amount of speech data. Setting (2) rises the number of clients to 100 making it a more challenging scenario as each device participates less frequently and has a smaller portion of data \cite{kairouz2021advances}. 
20 randomly selected clients participate in each round. 
\vspace{1mm}

\noindent \textbf{Federated SSL pretraining}. In each FL round, the SSL model is locally trained on each client with $1$ local epoch, batch size of $4$ in both FL settings. Then, the updated model weights are sent to the server for aggregation directly with FedAvg \cite{mcmahan2017communication} or after re-weighting the model updates using the client loss as in \cite{gao2021end}. The latter is better suited for heterogeneous data distributions across clients. 
The SSL model in 10-client setting, trained for $150$ rounds, shows a faster converging behaviour than 100-client FL which is trained for $500$ rounds (Fig. \ref{fig:loss} above). This is expected as a 100-client setting is more challenging. 
\vspace{1mm}

\noindent \textbf{ASR downstream evaluation}. The learned speech representations from SSL pretraining are evaluated with an ASR downstream task by adding $3$ linear layers on the top. The entire network is fine-tuned for $80$ epochs with CTC loss following the official SpeechBrain recipe. Figure \ref{fig:loss} (bottom) shows the WER on the test set.
Federated SSL representations provide competitive performance in both settings, demonstrating its feasibility while benefiting from unconstrained hardware environments (i.e., potentially unrealistic).
We show that loss-based aggregation achieves better performance in both settings, which indicates that weakening the effects of low-quality clients can assist the aggregation process in federated SSL.

\section{Conclusion}
In this paper, we provided the first empirical study of the feasibility of combining two emerging techniques that will revolutionize the way speech systems are designed and operate: SSL and FL. Our results preview the future in which hybrid FL and SSL systems train universal speech representations. Our study highlights this future will not be easy to reach, with a number of technical challenges ahead of us. But we hope this work will act as a roadmap to the community towards this new milestone.

%


\bibliographystyle{IEEEtran}

\bibliography{mybib}


\end{document}